\documentclass[aip,rsi,reprint,superscriptaddress,citeautoscript]{revtex4-1}
\usepackage{graphicx}
\usepackage{natbib}
\usepackage{floatrow}
\usepackage{amsmath}
\usepackage{array}
\usepackage{color}
\usepackage{hyperref} 
\usepackage[toc,page]{appendix}
\usepackage{xcolor}
\hypersetup{
  colorlinks   = true, 
  urlcolor     = black, 
  linkcolor    = blue, 
  citecolor   = blue 
}
\usepackage{leftidx}

\begin{document}

\title{High-resolution collision energy control through ion position modulation in atom-ion hybrid systems}
\author{Prateek Puri}
\affiliation{Department of Physics and Astronomy, University of California -- Los Angeles, Los Angeles, California, 90095, USA}
\email{prateek.puri01@gmail.com}
\author{Michael Mills}
\affiliation{Department of Physics and Astronomy, University of California -- Los Angeles, Los Angeles, California, 90095, USA}
\author{Elizabeth P. West}
\affiliation{Department of Physics and Astronomy, University of California -- Los Angeles, Los Angeles, California, 90095, USA}
\author{Christian Schneider}
\affiliation{Department of Physics and Astronomy, University of California -- Los Angeles, Los Angeles, California, 90095, USA}
\author{Eric R. Hudson}
\affiliation{Department of Physics and Astronomy, University of California -- Los Angeles, Los Angeles, California, 90095, USA}

\begin{abstract}
We demonstrate an ion shuttling technique for high-resolution control of atom-ion collision energy by translating an ion held within a radio-frequency trap through a magneto-optical atom trap. The technique is demonstrated both experimentally and through numerical simulations, with the experimental results indicating control of ion kinetic energies from $0.05-1$ K with a fractional resolution of $\sim10$ and the simulations demonstrating that kinetic energy control up to $120$ K with a maximum predicted resolution of $\sim100$ is possible, offering order-of-magnitude improvements over most alternative techniques. Lastly, we perform a proof-of-principle chemistry experiment using this technique and outline how the method may be refined in the future and applied to the study of molecular ion chemistry. 
\end{abstract}
\maketitle

\section{Introduction}

Reactant collision energy can strongly influence the kinetics and product outcomes of a reaction, revealing fundamental properties about the underlying chemical system~\cite{Bergeat2015,Klein2016,Gubbels2012}. Consequently, there has been much work on creating methods capable of precisely controlling this parameter. Lee, Herschbach, and coworkers developed the crossed molecular beam apparatus to explore the effect of collision energy on the angular distributions of products in neutral-neutral reactions, revolutionizing the field of gas-phase chemistry~\cite{Lee1969}. Other groups~\cite{Shagam2015,Henson2012,Amarasinghe2017} have since extended the technique to the millikelvin regime with improved energy resolution, enabling observations of quantum scattering resonances in reaction rates~\cite{Klein2016,Vogels2015,Kim2015,Dong2010}.
In this work, we take a step towards enabling similar high-resolution studies of ion-neutral reactions, which have been observed to play an important role in the formation of the interstellar medium and other astrophysical processes~\cite{Indriolo2010,Snow2008,Murray1999,Reddy2010}, by developing a novel technique for controlling collision energy in these systems.

Early efforts to control collision energies in ion-neutral systems, such as the SIFT~\cite{Smith1996,Armentrout2004} and CRESU~\cite{Rowe1987,Rowe1995} techniques, combined gas-discharge ion sources with neutral beams of tunable temperature. Current implementations of these experiments are typically restricted to collision energies of $\sim10-500$ K with fractional energy resolutions of $\sim10-100$~\cite{Smith2000}, depending on neutral beam parameters. We define the fractional resolution of a distribution $X$ as $R_X=\overline{X} / \sigma_{X}$ where $\overline{X}$ and $\sigma_{X}$ are the average and standard deviation of $X$, respectively. 

\begin{figure}[t]
\begin{center}
  \includegraphics[width=8.5cm]{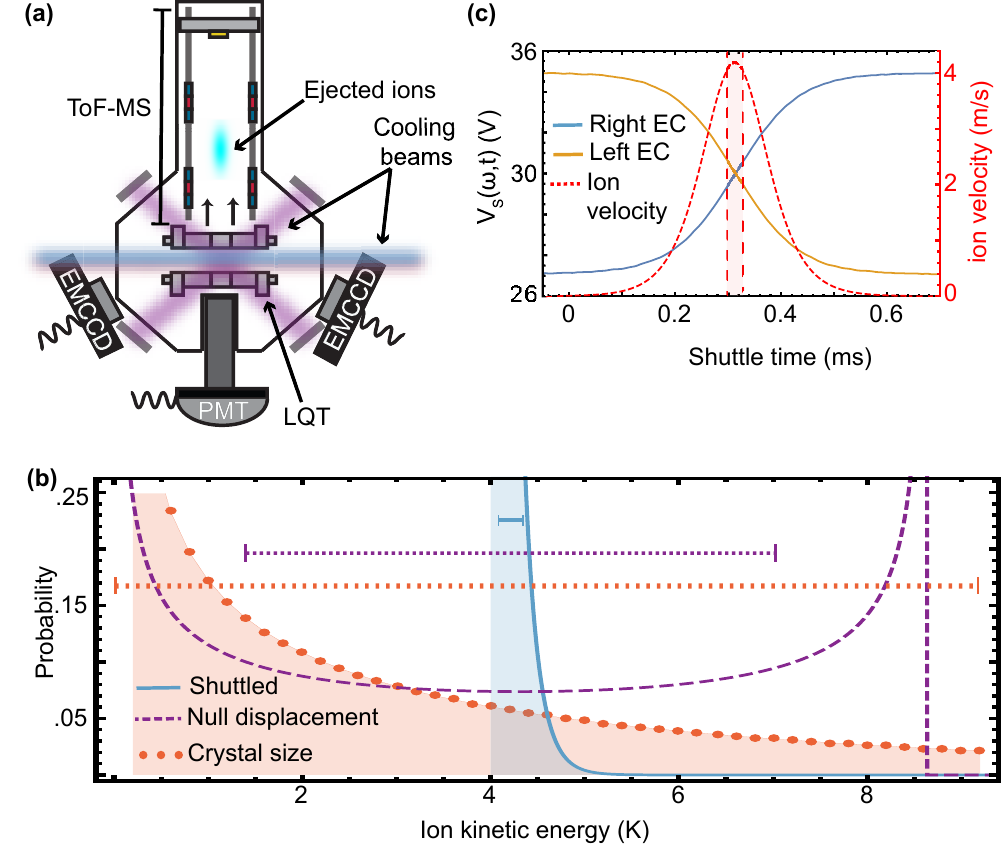}
\end{center}\vspace{-0mm}
\caption{ \textbf{Shuttling procedure and energy resolution} \\\hspace{\textwidth}
 \textbf{(a)} Schematic of the MOTion trap apparatus displaying an ion cloud being ejected from the $12$-segment LQT ($3$ segments per rod) into the ToF-MS, with arrows denoting the direction of ejection. \textbf{(b)} Energy distributions, derived from approximate Mathieu equation solutions, of a Yb$^+$ sample tuned to an average kinetic energy of $\sim 4$ K through ion chain displacement from the trap null, crystal size tuning, and idealized shuttling at a constant velocity of a crystal with an initial micromotion energy of $\sim 100$ mK. The standard deviations for each distribution are denoted by horizontal scale bars. \textbf{(c)} Voltage waveforms measured on the right and left endcap electrodes (EC) of the LQT, as well as the corresponding predicted ion velocities, expressed as a function of shuttle time. The waveforms follow the $V_S(\omega,t)$ profile, presented in Eq.~\ref{waveform}, with $V_{DC} = 30$ V, $V_{amp} = 5$ V, $\gamma = 0.18$, and $\omega = 2 \pi \cdot 95$ Hz. The portions of the waveform where the ions are stationary are not shown for clarity. The shaded region denotes the approximate period of overlap between the shuttled ions and the MOT.}
\label{Fig1}
\end{figure}

Ion-neutral reaction experiments have recently been extended to laser-cooled hybrid systems [Fig.~\ref{Fig1}(a)] capable of accessing millikelvin temperatures~\cite{Zhang2017,Zipkes2011,Harter2014}, where they have been used to explore reaction rate dependencies on conformational and electronic states~\cite{Chang2013,Sullivan2012,Hall2012,Rellergert2011,Ratschbacher2012} and to produce novel chemical species~\cite{Puri2017}. The majority of these hybrid systems contain radio-frequency (rf) ion traps, and typically the atom-ion collision energy is controlled by manipulating the ion excess micromotion energy, either by using electric fields to displace the ions from the rf trap null~\cite{Zipkes2010b,Zipkes2010,Hall2012} or by changing the size of the ion sample~\cite{Haze2013,Grier2009,Puri2017}. 

Upon displacing an ion from the rf trap null, the ion excess micromotion energy distribution approximately follows that of a simple harmonic oscillator [Fig.~\ref{Fig1}(b)]. The distribution stretches from $0$ K to twice its average value, with the average energy varying quadratically with radial displacement (all energies in this work are expressed in units $J/k_B=K$). Similarly, when using crystal size to tune the collision energy, each ion at a distinct radial position within the crystal has a unique harmonic distribution, and upon averaging over all radial positions, the resultant energy distribution is peaked at low energies with a high-energy tail. While the average kinetic energy of an ion sample can be precisely controlled using both of these techniques, their energy resolution is $\approx1$, making it difficult to measure energetically narrow features. Further, micromotion interruption collisions~\cite{Chen2014} and calculations of the atom-ion spatial overlap~\cite{Rellergert2012} may provide further complications for these micromotion-based techniques as collision energy is scanned. In particular, ions held within rf traps integrated into atom-ion hybrid systems are known to settle into Tsallis law energy distributions characterized by power-law tails after undergoing several collisions with an atomic sample~\cite{Chen2014,Rouse2017}, thereby leading to extreme high energy collision events that can jeopardize controlled collision energy studies. While active laser-cooling can mitigate many of these concerns, in certain cases they can still be nontrivial; however, the intricacies of such considerations will be omitted for simplicity in the following discussion.

Thus, other techniques have been developed to avoid these drawbacks and achieve higher energy resolutions. For example, Zeeman~\cite{Cremers2017,Dulitz2015} and Stark decelerators~\cite{Oldham2012} have been coupled to ion traps to probe atom-ion collision energies in the $\sim10-100$ K range with an energy resolution of $\sim50$~\cite{Parazzoli2009}. In other work, Eberle~\cite{Eberle2016} and coworkers recently demonstrated a novel method that uses optical ``push'' beams to precisely control the motion of atom clouds for kinetic energies ranging from $\approx 10-500$ mK with a resolution of$~\approx10$. 

Here, inspired by Eberle \textit{et. al.}~\cite{Eberle2016}, we describe a simple alternative that can be immediately used in most existing hybrid systems. In this technique, ions are translated at fixed velocities across a neutral sample by adjusting their axial trapping potential, maintaining the ions on the rf trap null throughout the process. At constant translational ion velocity, $R_E$ is primarily limited by the micromotion energy of the ion crystal [Fig.~\ref{Fig1}(b)] and can exceed values of $100$. In what follows, we describe the experimental system, investigate the shuttling technique through both experiment and simulation, and identify parameters where constant velocity ion motion can be approximately realized.  

\section{Experimental system}

The MOTion trap system employed in this study [Fig.~\ref{Fig1}(a)], described in detail elsewhere~\cite{Rellergert2012,Schowalter2016,Puri2017}, consists of a Ca magneto-optical trap (MOT) with a co-located linear quadrupole rf trap (LQT), which is in turn radially coupled to a time-of-flight mass spectrometer (ToF-MS)~\cite{Schowalter2012,Schneider2016}. An electron-mulitplying charged-coupled device (EMCCD) camera and a photomultiplier tube (PMT) capture fluorescence from the ions through a reentrant flange imaging system (NA$=0.2$), while the MOT atoms are imaged using two separate EMCCD cameras equipped with laser-line optical filters. 

An arbitrary waveform generator produces the endcap voltage waveforms that modulate the ion axial position. Output from the generator is amplified and low-pass filtered to remove any electrical noise near secular resonances of the trapped ions. 

\section{Shuttling principles}

For small displacements from the ion trap center, the electrostatic potential in the axial dimension $z$ at time $t$ is given as $U_{ax}(z,t) \approx \frac{\kappa V_{end}}{z_{A}^2} (z-z_0(t))^2$, where $\kappa$ is a factor associated with the ion trap geometry, $V_{end}$ is the endcap voltage, $z_{A}$ is the endcap electrode spacing of the LQT, and $z_{0}(t)$ is the time-dependent axial equilibrium position of the trap. In our system, $\kappa~\approx0.02$ and $z_A~\approx 10.2$ mm. By adding a time-dependent voltage waveform between right and left endcap electrodes [Fig.~\ref{Fig1}(c)], $z_{0}(t)$, and hence the ion crystal position, can be modulated at a speed proportional to the time derivative of the applied waveform. By changing the ramping speed of the waveform while keeping the peak-to-peak voltage constant, the translational velocity of the ion, and thus the ion kinetic energy $E$, can be conveniently controlled. 

When the modulation technique is used in conjunction with laser cooling, the motion of the resulting system can be described as a damped harmonic oscillator (DHO):

\begin{equation} \label{eq1}
m \ddot{z} = -k_{eff}(z,t)(z-z_0(t))+F_{\beta}(\dot{z}),
\end{equation}
where $m$ is the mass of the ion of interest, $k_{eff}(z,t) $ is the effective spring constant of the moving endcap potential, approximated as $q \frac{d^2}{dz^2} U_{ax}(z,t)$ where $q$ is the charge of the ion of interest, and $F_{\beta}(\dot{z})$ is a velocity-dependent damping force with an $e^{-1}$ motional damping time constant $\beta$ (units s$^{-1}$) determined by the laser parameters of a given Doppler-cooled system. Higher order terms are neglected. 

In order to achieve well-controlled energy resolution, the ion position should adiabatically follow the moving equilibrium position of the axial potential. However, if the Fourier transform of $z_0$(t) possesses frequency components near secular resonances of the ion, the shuttling motion may excite secular oscillations and heat the ion. To avoid this, we raise the trap axial confinement, thereby increasing the ion secular frequency above these frequency components, and strategically choose a ramping waveform less prone to ion heating.

\begin{figure}[!]
\begin{center}
  \includegraphics[width=8.5cm]{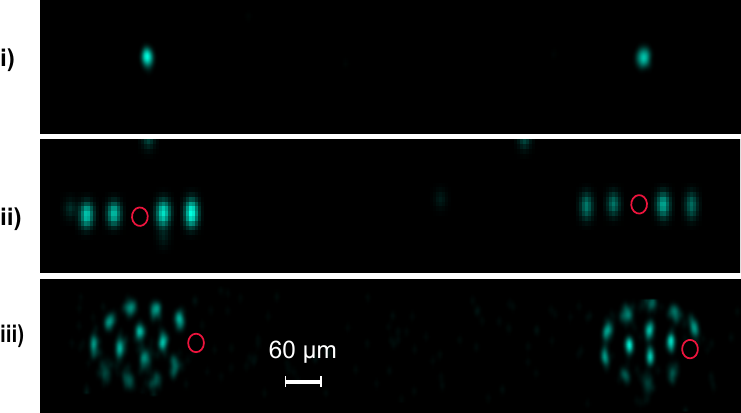}
\end{center}\vspace{-0mm}
\caption{ \textbf{Ion shuttling imaging} \\\hspace{\textwidth}
False-color experimental fluorescence images of Yb$^+$ ions undergoing shuttling presented for the following cases: i) a single ion, ii) a five-ion chain with one non-laser-cooled dark isotope, iii) a two-dimensional Coulomb crystal with one embedded dark isotope. As the ions spend $\sim 90\%$ of the time at the trajectory endpoints, their fluorescence is only evident in these locations. 
}
\label{Fig2I}
\end{figure}

Waveform optimization is a well-studied problem in the quantum information community~\cite{Hucul2007,Blakestad2009,Blakestad2011,Reichle2006}. In this work, when transporting an ion from one shuttling endpoint to another, we implement a hyperbolic tangent profile similar to that presented in Ref. \onlinecite{Hucul2007}, given by
\begin{equation} \label{eq2}
f_{tanh}(\tau) = \frac{\tanh{(2 \alpha \tau - \alpha)}}{\tanh{\alpha}} 
\end{equation}
where $\tau$ is the shuttle time as a fraction of total shuttle duration and $\alpha$ is a parameter that characterizes the slope of the function, chosen to have a value of $4$ in the work presented here. While a linear profile would seem to produce a flatter velocity profile at the trajectory midpoint, such a profile, once Fourier-decomposed, may possess frequency terms near secular resonances of the ion that could lead to secular heating, especially at high shuttle energies. The hyperbolic profile, while exhibiting a larger velocity spatial dependency, avoids these effects while still allowing for sufficient velocity control over the narrow region of MOT interaction such that other effects, such as excess micromotion compensation, are generally the limiting factor to energy resolution (see Section~\ref{singleIon}).

To meet the demands of our experiment, additional modifications were made to the applied shuttling waveform. Firstly, the waveform was chosen to be periodic in time to allow for waveform frequency, and thus ion velocity, to be varied while not affecting other experimental parameters, such as the time-averaged spatial overlap between the atom and ion sample~\cite{Sullivan2012}. Secondly, the waveform was constructed such that the ions remain at the stationary endpoints for a majority of the shuttling period, allowing sufficient laser cooling time to dampen any excitations that may occur during the transport process. A natural choice of waveform that satisfies the above criteria, shown in Fig.~\ref{Fig1}(c), is given by the following piecewise function:
\begin{multline}
\label{waveform}
V_{S}(\omega,t) = \\
\begin{cases} 
     V_{DC} - V_{amp} & 0 \leq t < \frac{T}{2}(1-\gamma) \\
     V_{DC} + f_{tanh}(\frac{t-\frac{T}{2}(1-\gamma)}{\frac{T}{2} \gamma}) V_{amp} & \frac{T}{2}(1-\gamma) \leq t < \frac{T}{2} \\
     V_{DC} + V_{amp} & \frac{T}{2} \leq t < \frac{T}{2}(2-\gamma) \\
		 V_{DC} - f_{tanh}(\frac{t-\frac{T}{2}(2-\gamma)}{{\frac{T}{2} \gamma}}) V_{amp} & \frac{T}{2}(2-\gamma) \leq t < T \\
   \end{cases}
\end{multline}
where $\omega$ is the angular shuttle frequency, $V_{DC}$ is the base endcap voltage, $V_{amp}$ is the amplitude of the shuttle waveform, $T=\frac{2 \pi}{\omega}$ is the shuttle period, and $\gamma$ is a factor that determines the ratio of stationary time to shuttled time during the ion trajectory. For a standard shuttle with an endpoint-to-endpoint distance of $\sim$$1$ mm, parameters are chosen as follows: $V_{DC}$ $\sim 30$ V, $V_{amp}$ $\sim 2$ V, $\gamma$ $\sim$ $0.1$, and $\omega$ can be tuned as desired from $\sim$ $2 \pi \cdot (0-500)$ Hz, providing control of the ion kinetic energy from $\approx0.01-10$ K. For reference, the axial secular frequency of our trap is typically chosen to be $\approx 2 \pi \cdot 30-150$ kHz for the range of axial confinements explored in this work. 

Fluorescence from the laser-cooled Yb$^+$ ions was collected with an EMCCD while shuttling. Shuttling images are presented in Fig.~\ref{Fig2I} for a single ion, a five-ion chain, and a two-dimensional Coulomb crystal, the last two of which are embedded with a non-laser-cooled Yb$^+$ isotope, indicating that this technique may also be used with sympathetically cooled species, such as molecular ions. 

\section{Experimental investigation of technique}
\label{exp}

\begin{figure}[!]
\begin{center}
  \includegraphics[width=8.5cm]{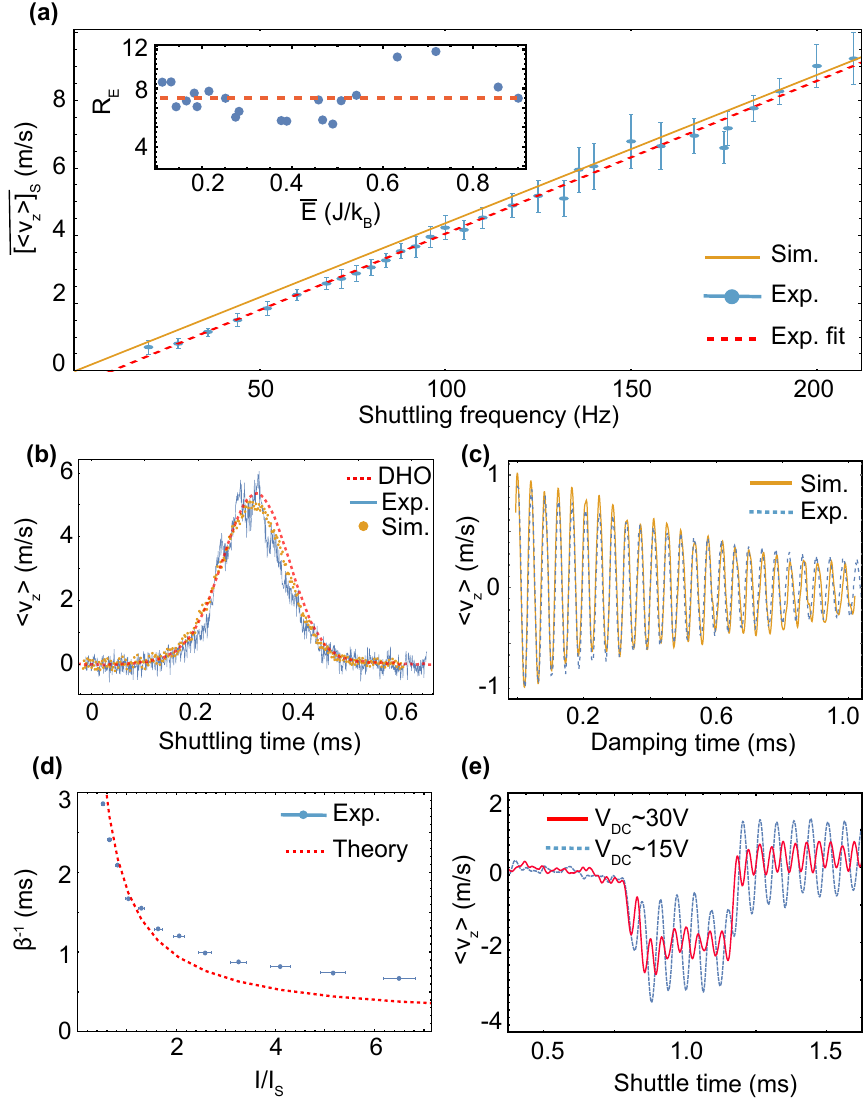}
\end{center}\vspace{-0mm}
\caption{ \textbf{Doppler velocimetry and large crystal simulation results} \\\hspace{\textwidth}
\textbf{(a)} Experimentally measured $\overline{[\langle v_z \rangle]_S}$ of a $\sim 100$ ion crystal obtained through Doppler velocimetry at different shuttling frequencies, where the error bars are displayed at the $1\sigma$ level. The experimental results show reasonable agreement with MD simulations. A linear fit applied to the experimental data shows that varying the shuttle frequency modifies the axial velocity of the trapped ions in the expected way. The inset shows the corresponding mean kinetic energies and energy resolutions obtained at the various shuttling frequencies, with the dotted line referring to the average resolution. Note here that the plot refers to averages and resolutions of the distribution $[\langle \hat{E_z} \rangle]_S$ (see text), but the subscript was omitted in the plot for clarity. \textbf{(b)} Experimental $\langle v_z \rangle$ values, obtained as a function of shuttle time at a shuttle frequency of $120$ Hz, are compared to results of a MD simulation and the predictions of a $1$D damped harmonic oscillator model. \textbf{(c)} The effects of laser cooling on damping secular motion from both simulation and experiment. The saturation parameter used to construct the laser cooling force in the simulations was tuned until $\beta$ matched well with experiment. \textbf{(d)} Experimental damping timescales are obtained as a function of laser cooling saturation parameter and are compared with predictions from a rate equation model. Horizontal and vertical error bars are expressed at the $1\sigma$ level, with the latter being smaller than the data points \textbf{(e)} Measured $\langle v_z \rangle$ as function of shuttle time for two different axial confinement strengths. The shuttle was performed with a linear ramping profile more prone to ion heating than $f_{tanh}(t)$ in order to accentuate the increase in energy resolution that is possible with greater axial confinement.}
\label{Fig2}
\end{figure}

\subsection{Fluorescence detection while shuttling} \label{flor}

Understanding the energy dynamics of the shuttled ions requires knowledge of their velocity distribution. In order to experimentally characterize this distribution, a $^{174}$Yb$^+$ crystal was shuttled over a $\approx1$ mm distance by applying $V_S(\omega,t)$ to the endcap electrodes of the ion trap. The EMCCD in our imaging system was replaced with a PMT to record the ion photon scattering rate throughout the shuttling process.

The photon scattering rate of laser-cooled Yb$^+$ can be approximated~\cite{Metcalf2003} as

\begin{equation} \label{eq3}
\Gamma_{scatt} (v_z) = \frac{\Gamma}{2}\frac{s}{1+ s + 4\frac{(\delta-k_z v_z)^2}{\Gamma^2}}
\end{equation}
where $\Gamma$ is the transition linewidth, $s$ is the saturation parameter of the cooling laser, given as $I/I_s$ where $I$ is the intensity of the laser beam and $I_s$ is the saturation intensity of the transition, $\delta$ is the detuning of the laser from resonance, $k_z$ is the magnitude of the k-vector of the axially-aligned cooling laser, and $v_z$ is the z-component of the ion velocity. The scattering rate is insensitive to micromotion or secular motion in the radial dimension. 

\subsection{Ensemble and spatial averaging of velocity distributions}

For a single shuttled ion, the velocity-dependent scattering rate (see Eq.~\ref{eq3}) during each PMT time acquisition bin ($\sim10$ ns width) can be used to measure $v_z(t)$. In order to enhance the PMT signal in the experiment, we interrogate an ensemble of $\sim100$ ions, and consequently, the velocities extracted in each time bin are ensemble averages of the total axial velocity distribution, defined as $\langle v_z \rangle$ (Appendix.~\ref{Doppler}). Here, we define the ensemble average of a data set $X$ as $\langle X \rangle$.

Further, the shuttled ions are only overlapped with the neutral sample at the center of the ion trap for a small portion of their trajectory. Therefore, the relevant resolution to consider is the resolution of $[\langle v_z \rangle]_S$, the velocity distribution with weighting factors determined by the spatial overlap of the ions at each shuttle time with an atom sample of characteristic length scale $w_A$ (Appendix.~\ref{spatial}). Here, we define the spatially-weighted distribution of a data set $X$ as $[X]_S$~\cite{Patil2014}. For optimal resolution, the neutral sample should be placed at the center of the ion trajectory where the ion velocity is most constant. 


\subsection{Simulation parameters} \label{sim}

The experimental results are compared to predictions of molecular dynamics (MD) simulations conducted with the SIMION $8.1$ software package~\cite{Dahl2000}, as shown in Fig.~\ref{Fig2}(a) and~\ref{Fig2}(b). The simulation software employs finite difference methods to numerically solve Laplace's equation for a given set of electrodes and point charges, allowing for determination of ion trajectories and energy distributions. Time-dependent trapping potentials were incorporated into the simulation to properly include the effects of micromotion, and ion-ion repulsion was treated by superimposing the Coloumb interaction from co-trapped ions with the potential produced by the quadrupole trap electrodes. The simulations were performed using $100$ ions, approximately equivalent to the number used during the experiment, and also employed a laser-cooling damping force whose velocity profile was derived from a simple four-level rate equation model. 

In order to optimize the accuracy of the simulated laser cooling force, both in experiment and simulation, the ions were initialized in the LQT, non-adiabatically transported between trajectory endpoints through a square-wave-like voltage ramp, and subsequently observed as the laser cooling force damped the motion of the excitation [Fig.~\ref{Fig2}(c)]. The saturation parameter of the simulated laser cooling force was adjusted until the $e^{-1}$ decay constant $\beta$ matched that observed in experiment. We also investigated how this damping timescale varied with laser cooling intensity by repeating the above measurement at various laser powers [Fig.~\ref{Fig2}(d)]. The results are comparable with those expected from our rate equation cooling model and are instructive when considering what laser cooling parameters to implement while shuttling. Namely, one should operate in a laser cooling regime such that the time spent at the shuttle endpoints during each cycle is much longer than the damping time, ensuring the ions are sufficiently cooled before the next shuttle cycle begins. Further, the simulations confirm that at experimental conditions, the laser cooling damping force does not significantly influence the trajectory of the ions while shuttling.

\subsection{Analysis of results} 

The experimental and simulated results for $\overline{[\langle v_z \rangle]_S}$ are in reasonable agreement [Fig.~\ref{Fig2}(a)]. Both exhibit a linear relationship with waveform frequency, affirming that $w_A$ can be varied to predictably control the velocity, and thus collision energy, of the ions. The trajectory for the ions assuming a damped harmonic oscillator model, shown in Fig.~\ref{Fig2}(b), also appears to describe the ion motion well, confirming that the model may be used to gain intuition about the shuttling procedure. We attribute minor discrepancies between the simulation and experiment, such as differing damping timescales and amplitudes of secular oscillation while shuttling, to imperfect voltage matching due to unmeasured electrode charging and rf pickup, minor discrepancies in laser cooling velocity profiles, and effects not considered in the simulation such as micromotion interruption collisions with background gas particles. 

The experimental energy resolution can also be compared to predictions from simulation. $\overline{[\langle \hat{E_z} \rangle]_S}$, defined as $\frac{1}{2} m (\overline{[\langle v_z \rangle^2]_S})$, was scanned over $\approx0.01-1$ K over the velocities explored in Fig.~\ref{Fig2}(a), with probing of higher kinetic energies precluded by difficulty in discriminating between scattering rates at large $v_z$. Shown in the inset to Fig.~\ref{Fig2}(a), the measured ensemble-averaged axial energy resolutions, $R_{[\langle \hat{E_z} \rangle]_S}$, were determined to be $\approx 10$, in agreement with simulations.

However, the resolution of the non-ensemble-averaged kinetic energy distribution, $[E_{z}]_S=\frac{1}{2} m [v_z^2]_S$, is the more relevant quantity to consider when characterizing collision energy control since it is sensitive to center-of-mass frame velocity dispersions. Measuring $R_{[E_{z}]_S}$ involves knowing the velocities of each individual ion, information unavailable with our velocimetry technique. Therefore, we utilize the simulations to estimate this quantity and obtain $R_{[E_{z}]_S} \approx6$ (Appendix.~\ref{EnRes}).

Experimental average velocity distributions were also obtained at various levels of axial confinement, and, as expected, higher axial confinement offered superior resolution. To exaggerate this effect, we performed a shuttle using a linear ramping profile prone to ion-heating and observed that increased confinement more effectively suppressed secular oscillations [Fig.~\ref{Fig2}(e)]. Probing of even higher axial confinements was prohibited in our system by technical considerations. 

\section{Single ion and molecular ion simulation results} \label{singleIon}

\begin{figure}[!]
  \includegraphics[width=8.5cm]{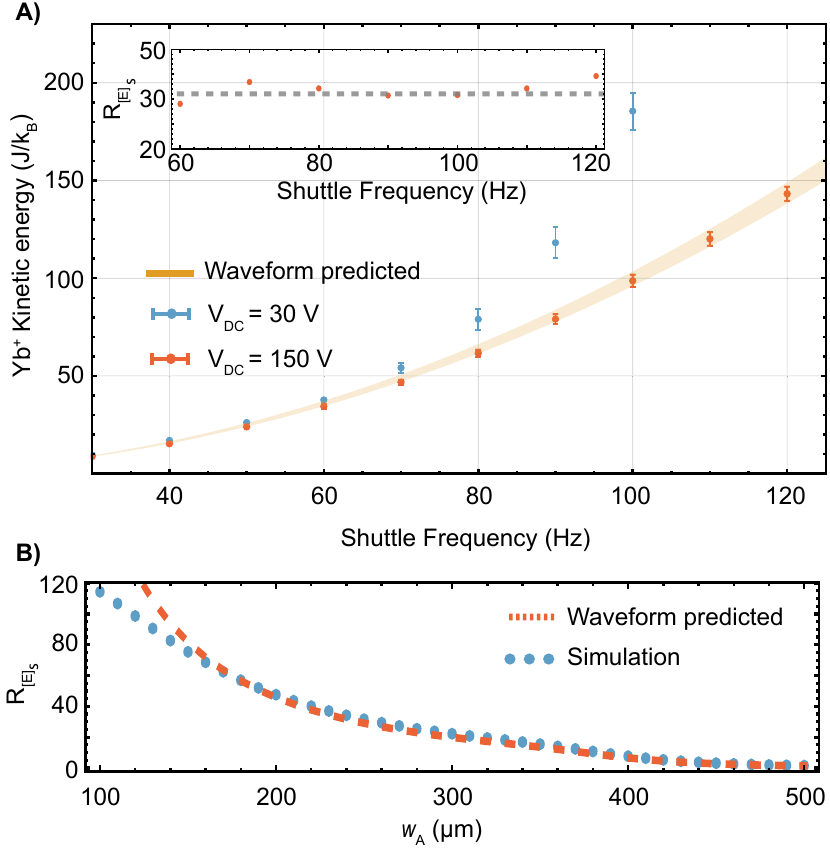}
\caption{\textbf{Single ion simulation results} \\\hspace{\textwidth}
\textbf{(a)} Total kinetic energy for a simulated single ion shuttled at various waveform frequencies, using two separate axial confinements. The simulations performed at the higher axial confinement display higher energy resolutions and exhibit less significant secular oscillations, as evidenced by their adherence to the waveform-predicted energy, shown in bands. Error bars are expressed at the $1\sigma$ level. The inset to the figure shows $R_{[E]_S}$, the total kinetic energy resolution including both axial and radial motion, for the high axial confinement simulations, with the average resolution of $\approx 35$ denoted by the dotted line. Low axial confinement simulations produced average energy resolutions of $\approx 20$. \textbf{(b)} $R_{[E]_S}$ of a simulated ion shuttled at $\approx 100$ K of kinetic energy as a function of neutral cloud spatial dimension. The results are compared to the resolutions that would be expected if the ions perfectly followed the motion of the equilibrium position without any micromotion or secular excitation.}
\label{Fig3}
\end{figure}

While Section~\ref{exp} demonstrates large ion samples can be successfully shuttled, the resolution is maximized when used with a single ion, where ion heating effects are minimal and the ion shuttling energy can dominate over its micromotion energy. In this section, kinetic energy will refer to the total kinetic energy of the ion, including both axial and radial motion. 

While our experimental optical detection efficiency prevents extensive single ion measurements, reasonable best-case-scenario simulations are performed with a single Yb$^+$ ion utilizing the electrode geometry of the MOTion trap and the laser cooling profile described in Section~\ref{exp}. Further, laser cooling while shuttling was necessary in the work discussed in Section~\ref{flor} for Doppler velocimetry purposes, but in general laser cooling may be switched off during transport if, for example, finer control of ion electronic state populations is desired. However, we choose to maintain laser cooling throughout the shuttling process in the following simulations for consistency with the simulations performed in the previous section. The end-to-end shuttle distance, experimentally limited to $\approx 1$ mm by the field of view of our imaging system, is increased to $\approx 2$ mm to enhance energy resolution. Further, idealized waveforms were implemented in the simulation instead of the waveforms measured in the experiment, where unintended filtering due to trap electronics and rf pickup caused slight waveform distortion. 

These simulations, presented in Fig.~\ref{Fig3}(a), were performed at two different axial confinements ($V_{DC} = 150$ V and $35$ V) and once again indicate that confinement plays a pivotal role in determining energy resolution. When the endcap voltages were raised to $150$ V in the simulation, the rf voltage amplitude was also increased by a factor of $2$ relative to the low-confinement case to prevent radial defocusing caused by the increased axial confinement. At high enough axial confinements, further resolution improvement is eventually limited by the need to operate at increasingly large rf voltages to avoid this defocusing effect, forcing the trap towards high Mathieu $q$-parameter regimes where the ions become unstable. 

At higher energy in particular, the results presented in Fig.~\ref{Fig3}(a) indicate that low axial confinement can facilitate secular excitation of the ion motion. The excitation can cause the ion to either lag or lead the equilibrium position of the moving potential during MOT interaction, increasing the ion kinetic energy spread. At high enough energies, the shuttling process is no longer adiabatic, leading to large-scale secular oscillations that significantly broaden the ion kinetic energy distribution, as evidenced in the increasing energy spreads for the low axial confinement points in Fig.~\ref{Fig3}(a). However, increasing the axial confinement postpones this behavior until higher energies. For $V_{DC} = 150$ V, the ion position follows $z_0(t)$ closely for kinetic energies up to $120$ K while the kinetic energy resolution, $R_{[E]_S}$, approaches $35$ for $E \gtrsim 2$ K. Here the resolution is limited by a combination of increased micromotion energy at the large axial confinement, minor secular excitation during transport, and non-uniformities in the velocity profile of the shuttling waveform.

On the other hand, for $E \lesssim 2$ K where secular oscillations play less of a role, the resolution is ultimately limited by excess micromotion compensation techniques, which are typically accurate to within $\sim 10$ mK in quadrupole traps with dimensions similar to that used in this work. In this low energy regime, $V_{DC} \approx 5$ V is optimal since the reduced confinement limits micromotion from radial defocusing, permitting $R_{[E]_S} \approx20$. 

While a $250$ $\mu$m neutral cloud size was assumed when computing the energy resolutions in Fig.~\ref{Fig3}(a), further resolution increases can be realized by reducing the size of the neutral atom sample, thereby also reducing the sampled velocity spread of the ion trajectory. Often the spatial dimensions of neutral atom traps can be conveniently tuned using optical or magnetic fields, with some atom systems, such as dipole traps, approaching $5$ $\mu$m in size~\cite{Zuo2009}. In Fig.~\ref{Fig3}(b), $R_E$ is shown as function of $w_{A}$, with resolutions in excess of $100$ predicted for atom traps nearing the $100$ $\mu$m regime. Conversely, resolution may also be improved by increasing the distance between shuttling endpoints for a fixed atom cloud size. Increases in shuttle distance would also have the added benefit of mitigating secular oscillations as a lower frequency waveform with Fourier components further spaced from ion secular resonances could be used to obtain a given shuttle velocity. However, this improvement would come at the expense of more difficult micromotion compensation, as to be discussed below.

The simulations do not consider the effect of atom-ion collisions on the ion trajectory; however, at experimental atomic densities ($\approx 10^{10}$ cm$^{-3}$), over the range of energies explored in the simulations, there is a $\approx10^{-3}$ probability of a collision occurring with the MOT atoms in a given shuttle cycle. Therefore, any deviations from the expected ion motion caused by collision events are not expected to influence the energy of subsequent collisions, as there is only a $\approx10^{-6}$ probability of a second collision occurring before the ion motion is reinitialized through laser cooling at the trajectory endpoints. Additionally, to reduce the effect of background gas collisions on the ion trajectory, the technique may be used in ultra-high vacuum conditions. 

Further, the technique may ultimately be limited by effects unconsidered in the simulations, such as patch potentials and electrode charging, that make it difficult to optimally micromotion compensate at each trajectory position, especially given the comparatively large size of the utilized ion trap and the limited number of compensation electrodes. 

For example, in our system, if excess micromotion compensation is performed at the center of the shuttling trajectory, we experimentally observe $\sim100$ mK of uncompensated excess micromotion at the trajectory endpoints $2$ mm  displaced from the center point. While proper compensation throughout the trajectory may be a challenge in certain applications, we note that proper compensation in the narrow region of MOT interaction is most important for determining collision energy resolution, as the micromotion amplitude of the ion motion generally adiabatically follows any local uncompensated electric field (see Appendix~\ref{micro} for a more detailed treatment on the effects of excess micromotion on the shuttling process). Further, axial micromotion may provide additional complications, although radial micromotion will likely dominate this effect. Through simulations performed using our system, we observe less than a $< 2$ mK difference in ion energy due to axial micromotion between the center of our shuttling trajectory and a point displaced $2$ mm from the center; however, experimental imperfections may further increase this value.

To minimize these effects, the appropriate electrode shim voltages can first be identified for the ions at each trajectory position while the ions are stationary. Subsequently, the shim voltages can be updated while shuttling to ensure uniform micromotion compensation as the ion transits from one endpoint to the other. Additionally, excess micromotion compensation techniques, such as photon cross-correlation spectroscopy~\cite{Berkeland1998} or parametric excitation~\cite{Keller2015}, may be used to compensate micromotion with greater precision and maintain ions with excess micromotion energies nearing $\lesssim 5$ mK.

While the precise kinetic energy control of laser cooled species is beneficial, ultimately this technique may be most useful when applied to molecular ion chemistry, where it can be used to detect nuances in long range capture models~\cite{Song2016} and possibly illuminate rotational and vibrational resonance features that have thus far evaded current techniques. To explore this possibility, simulations are performed while shuttling two laser cooled Ba$^+$ ions and a sympathetically cooled BaCl$^+$ molecular ion, with the resulting energy distributions of the molecular ion depicted as a function of shuttle frequency and trajectory position in Fig.~\ref{Fig4}(a) and Fig.~\ref{Fig4}(b), respectively. In contrast to Yb$^+$, Ba$^+$ possesses a $\Lambda$ level-structure system, and thus, the three-level optical Bloch equations are solved to account for coherent-population-trapping effects in the simulated laser cooling force. 

The results from the simulation demonstrate that, similar to the Yb$^+$ single-ion case, energy resolutions for BaCl$^+$ approaching $40$ are achievable assuming a neutral atom cloud size of $250$ $\mu$m, a value over a order-of-magnitude greater than that offered by alternative micromotion-based techniques in this energy range and one that can be further improved by changing the axial confinement and employing a smaller atom cloud size, as discussed earlier in this section. At kinetic energies below $1$ K and when combined with a light mass atomic partner that would yield a low reduced mass, this resolution may be sufficient to resolve reaction resonance features, which have been predicted to have collision energy widths of order $\sim1-10$ mK~\cite{Silva2015}, although the particulars of the resonance of interest and control of the systematics alluded to above will ultimately determine if this is feasible. Here, the collision energy is proportional to the reduced mass of the atom-ion system, and in most current hybrid systems the average and the width of its distribution are typically a factor of $\approx1-10$ smaller than the corresponding kinetic energy values.

\section{Demonstration of technique for charge exchange reaction investigation}

\begin{figure}[!]
\begin{center}
  \includegraphics[width=8.5cm]{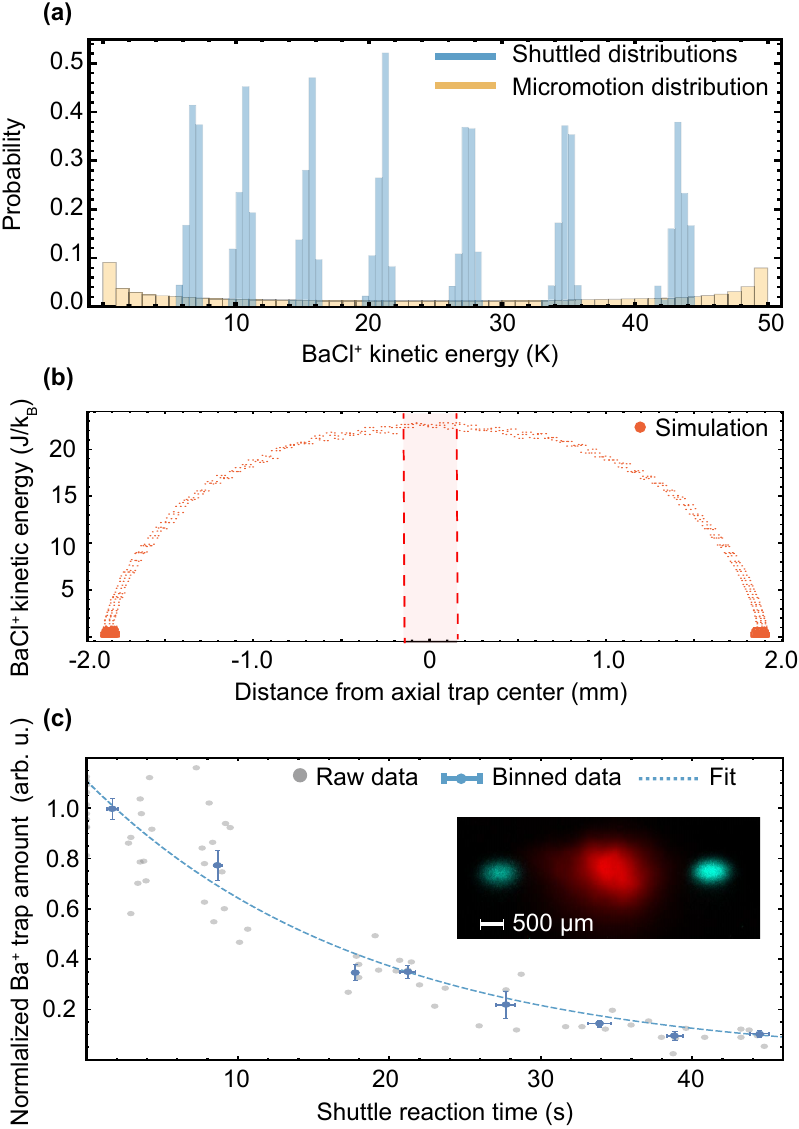}
\end{center}\vspace{-0mm}
\caption{ \textbf{Controlled chemistry implemented with ion shuttling} \\\hspace{\textwidth}
\textbf{(a)} Simulated energy distributions for a single shuttled BaCl$^+$ molecular ion sympathetically cooled by two laser-cooled Ba$^+$ ions. The shuttled distributions are presented for a variety of shuttle frequencies and are compared to the theoretical distribution obtained from using the excess micromotion of a single ion to access an average kinetic energy of $\approx 25$ K. \textbf{(b)} Simulated BaCl$^+$ kinetic energy as a function of axial ion position while shuttling. The dashed lines enclose the $250$ $\mu$m effective region of MOT interaction where the ion velocity is approximately constant. \textbf{(c)} Decay of Ba$^+$ amount from the LQT as a function of shuttling time when a Ca MOT is placed at the center of the trajectory. The inset displays superimposed experimental fluorescence images of a $\sim 500$-ion Ba$^+$ sample and a Ca MOT containing roughly one million atoms taken while performing a shuttling reaction rate measurement. The large ion sample utilized in the experiment was initially liquid upon loading into the LQT and remained so while shuttling.} 
\label{Fig4}
\end{figure}

As a proof-of-principle experiment, a cloud of $\approx 500$ Ba$^+$ ions was loaded into the LQT and shuttled through a Ca MOT located at the center of the ion trajectory at an average kinetic energy of $14(4)$ K. Here the resolution was limited by the inherent excess micromotion energy of the three-dimensional crystal. Fig.~\ref{Fig4}(c) shows the decay of Ba$^+$ ion amount in the LQT, measured by the ToF-MS~\cite{Puri2017}, as a function of shuttling duration due to charge-exchange collisions with ground-state Ca. The inset to Fig.~\ref{Fig4}(c) presents superimposed images of the atoms and ions obtained during shuttling, with each image taken using separate laser line optical filters. The geometric overlap between the atoms and ions was measured by phase-triggering the EMCCD cameras on the shuttling waveform to acquire ion images, and hence ion positions, at various points along their trajectory. This technique allowed for the effective imaging of ions with velocities $\lesssim 50$ m/s, bounded by effects related to the minimum camera exposure time of $10$ $\mu$s. For velocities in this range, the ions are seen to follow the expected shuttling trajectory, and numerical simulations are used to verify this trend at higher collision energies. After the overlap factor was verified, measured atomic densities were used, in a manner similar to Ref.~\onlinecite{Puri2017}, to calculate a total reaction rate of $2.4 (4) \times 10^{-11}$ cm$^{3}$s$^{-1}$, a value consistent with a previously measured result~\cite{Sullivan2012} after ion excited state fraction normalization. 

This proof-of-principle experiment demonstrates that this technique can be used to measure accurate rate constants for reactions between laser-cooled species and neutrals, paving the way for similar studies incorporating sympathetically cooled molecular ions. 

\section{Conclusion}

Blending techniques from the quantum information and hybrid trapping communities, we have demonstrated a method for controlling ion-neutral collision energy based on ion axial position modulation that is capable of offering energy resolutions, $\overline{E}/\sigma_E$, from $\sim 10-100$ over kinetic energies ranging from $\approx 0.05$ $-$ $120$ K. This combination of both range and resolution improves on alternative techniques that typically compromise one for the other. In addition to investigating the technique through experiment and simulation, we also performed a reaction rate measurement by shuttling laser-cooled atomic ions, and we suggested how the shuttling method may be implemented in future experiments to study molecular ion chemistry.

Further, the technique can be immediately implemented in currently existing hybrid traps with little experimental overhead. The shuttling procedure is also quite adaptable, and properties such as axial confinement, neutral atom size, and endpoint-to-endpoint shuttle distance can be custom-tailored to a variety of experimental conditions to obtain desired energy resolutions while obeying most experiment-specific constraints. 

Additional improvements may further increase the effectiveness of the technique. Ion traps with mulitple-segmented endcap electrodes that can more reliably compensate micromotion and produce more pure harmonic potentials throughout the trap may be utilized, allowing the ions to be shuttled over longer axial distances and while minimizing their acceleration in the MOT region. Further, if laser cooling during transport is necessary for a particular application, Doppler shifting of the ions while shuttling may be problematic if constant electronic state populations are desired. To this end, one may choose to appropriately adjust the frequency of the Doppler cooling laser while shuttling in order to produce a constant effective laser detuning. In addition, a imaging system with higher capture efficiency and a radial probe beam may be used to apply the Doppler velocimetry technique towards detecting the radial micromotion of single ions while shuttling and thus set more realistic bounds on excess micromotion compensation. 

Lastly, the waveforms utilized in this proof-of-principle study were largely chosen out of convenience and speed of implementation. While sufficient for the purposes of this work, they are by no means optimal. More sophisticated waveforms~\cite{Oswald2015,Furst2014,Clercq2016} that maintain flatter velocity profiles while not inducing secular heating may be used if even finer energy resolution is required.

\section{Acknowledgments}

This material is based upon work supported by the National Science Foundation Graduate Research Fellowship under Grant No. DGE-1650604. This work was also supported by the National Science Foundation (Grant No. PHY-1255526) and Army Research Office (Grants No. W911NF-15-1-0121 and No. W911NF-14-1-0378). 

\section{References}
%

\bibliographystyle{aipnum4-1_nourl}	

\appendix

\section{Doppler-cooling velocimetry} \label{Doppler}

In order to extract experimental velocities for shuttled ion samples, we first take a linescan of a sample of stationary ions where ion fluorescence is collected by the PMT at various cooling beam detunings, allowing for determination of both $s$ and $\delta$ in Eq.~\ref{eq3} in the manuscript. Laser cooling detunings are determined from a calibrated High Finesse wavemeter coupled directly to the laser beams in our system. 

Once these two parameters have been determined, the ions can then be shuttled while their total photon scattering rate is measured by the PMT during each acquisition time bin ($\sim 10$ ns). For a single ion, after normalizing by the stationary ion count rate, a time-dependent ratio, $\eta$(t), is produced which can be used to solve the equation $\eta(t) = \frac{\Gamma_{scatt} (v_z(t))}{\Gamma_{scatt} (v_z=0)}$ for $v_z(t)$. This velocimetry technique is only effective at determining motion along the axial propagation direction of our cooling beam and is insensitive to micromotion or secular motion in the radial dimension. 

Furthermore, since the stationary count rate of our sample changes over the course of the experiment, due to ion depletion caused by background gas reactions or slight laser power fluctuations, we renormalize our background ion count rate by collecting ion fluorescence from the stationary endpoints of the trajectory during each shuttling procedure. We also collect a stationary ion count rate at each point along the shuttle trajectory to ensure that ion fluorescence changes are indeed caused by velocity changes and are not a result of differing laser cooling alignment or light collection efficiency along the trajectory. 

Due to the low optical detection efficiency of the system, an ensemble of ions must be interrogated in order to produce a measurement with an adequate signal-to-noise ratio on reasonable experimental timescales. Consequently, the resulting PMT signal yields the collective sum of photons captured from the entire crystal of ions, making individual ion motion indiscernible. Our velocimetry technique therefore effectively measures $\langle \eta (t) \rangle = \frac{ \sum_i \Gamma_{scatt} (v^i_z)} {N \Gamma_{scatt} (v_z=0)} \approx \frac{\Gamma_{scatt} (\langle v_z \rangle)}{\Gamma_{scatt} (v_z=0)}$, where $\langle \eta (t) \rangle$ is the ensemble average count ratio, $N$ is the total number of ions in the system, $v^i_z$ is the axial component velocity of the $i^{th}$ ion, and $\langle v_z \rangle$ is the ensemble average axial velocity of the entire system. The latter approximation is justified in our regime since the dispersion of velocities is expected to be low during the shuttling process, as evidenced by simulations described in Section~\ref{exp}. 

\section{Spatial and ensemble averaging of distributions} \label{spatial}

The velocity and also the energy of the shuttled ions exists as a distribution in two separate dimensions. Firstly, for a multi-ion sample, at a given instance of time, the ions reside in a distribution of velocities within the crystal itself. The ensemble-averaged measurements extract the mean of this distribution while remaining insensitive to its spread, which is determined by ion-ion collision events and differences in excess micromotion energy. Secondly, this distribution of velocities within the crystal also changes as a function of shuttle time as the ions evolve along their trajectory, meaning all ensemble-averaged measurements will change as well. To assign weighting factors to each measurement, we consider the spatial overlap of the ions at each instance of time with a neutral sample located at the center of the trajectory. 

The neutral atom density distribution is approximately Gaussian. The spatial weighting factor associated with the velocity or energy measured in each time bin along the trajectory is defined as $w_b = \frac{e^{-(z_b/w_{A})^2}}{\lambda} $ where $z_b$ is the position of the ion during the $b_{th}$ time bin, $w_{A}$ is the $e^{-1}$ decay length scale of the atomic density distribution, and $\lambda$ is a normalization factor chosen such that $\sum_b w_b=1$. After the weighting factors have been calculated, the mean and standard deviations of the weighted distribution can be computed to yield the relevant distribution resolutions~\cite{Patil2014}.

\section{Experimental energy resolution} \label{EnRes}

We note the distinction between the energy distributions $[\langle \hat{E_z} \rangle]_S$ and $[E_{z}]_S$. $[E_{z}]_S=\frac{1}{2} m [v_z^2]_S$ while $[\langle \hat{E_z} \rangle]_S=\frac{1}{2} m ([\langle v_z \rangle^2]_S)$. $[\langle \hat{E_z} \rangle]_S$ only approximately describes the average kinetic energy of the sample as it assumes this quantity is proportional to $[\langle v_z \rangle^2]_S$ instead of $[\langle v_z^2 \rangle]_S$, the latter of which is incapable of being measured in experiment. However, this approximation is reasonable if the energy dispersion of the sample is expected to be small, as suggested by simulations. 

Further, $[\langle \hat{E_z} \rangle]_S$ is a distribution of ensemble-averaged energies and is distinct from $[E_z]_S$, the distribution of the non-ensemble-averaged axial kinetic energies. The latter is the more relevant distribution to consider when characterizing overall atom-ion collision energy control as it contains information on the spread of the entire axial energy distribution and not just the spread in the average energy of the sample. The simulations predict $R_{[E_z]_S} \approx 6$, while $R_{[\langle \hat{E_z} \rangle]_S}\approx10$ was measured from experiment. $R_{[E_z]_S}$ is smaller than $R_{[\langle \hat{E_z} \rangle]_S}$ for a variety of reasons. For example, within a large ion crystal, the ions may experience slightly different potentials at different points within the crystal and therefore reach different peak velocities, dispersing their overall energy distribution while keeping their average energy constant.

\section{The effect of micromotion on shuttling trajectories} \label{micro}

\begin{figure}[!]
  \includegraphics[width=8.5cm]{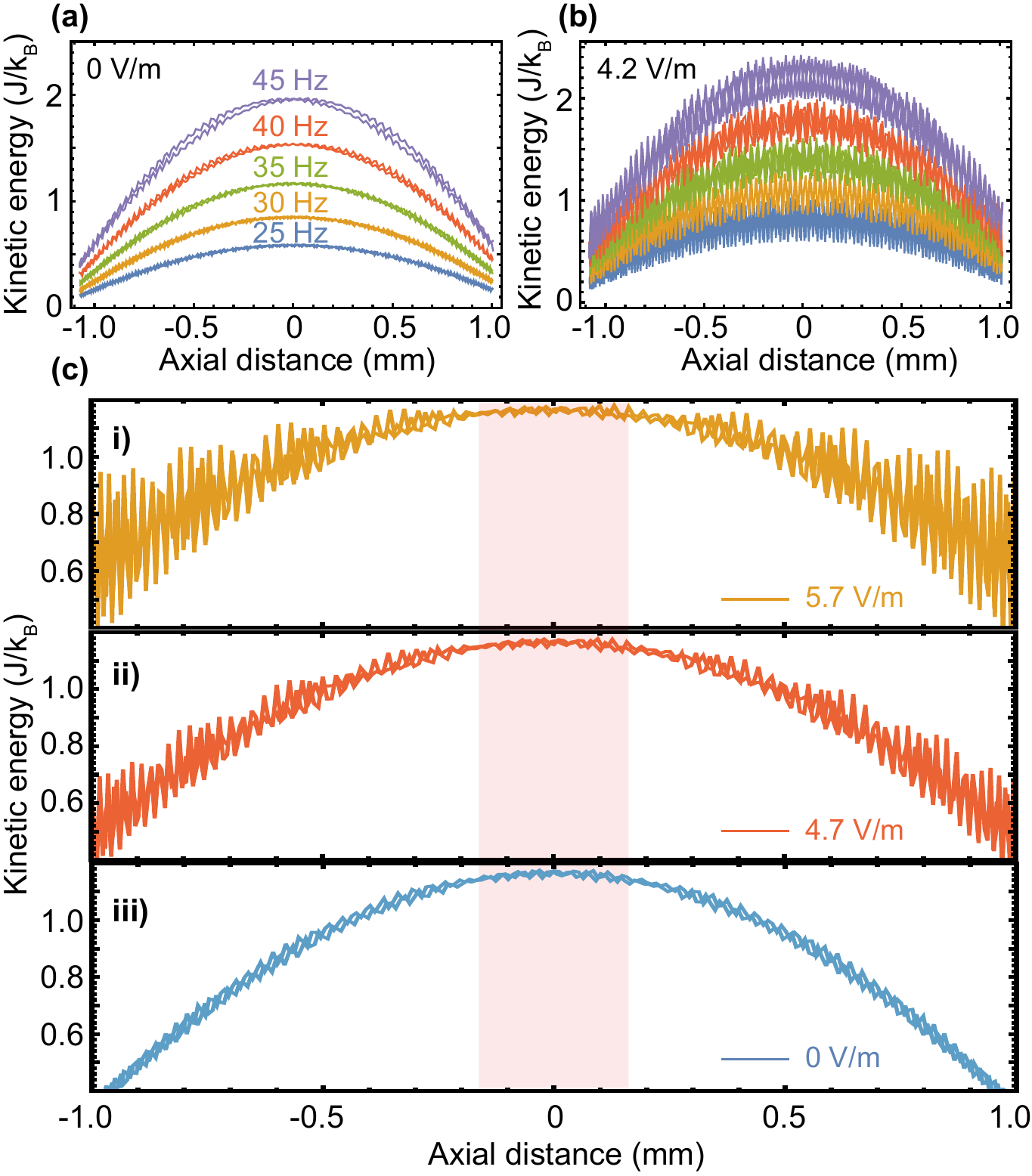}
\caption{\textbf{Effects of micromotion on shuttling energy} \\\hspace{\textwidth}
\textbf{(a)} Energy as a function of axial shuttle distance for a single ion in a perfectly excess micromotion compensated system at various shuttling frequencies. \textbf{(b)} The same simulations in (a) repeated with an additional electric field offset of $4.2$ V/m. The micromotion broadens the energy distribution and shifts the average energy upwards from the perfectly compensated simulations in (a). The $4.2$ V/m offset is consistent with what can currently be compensated in our system.\textbf{(c)} Simulations of an ion being shuttled with radial electric field gradients of varying strengths \textbf{(i-iii)} imposed upon the ion. The ions are assumed to be perfectly compensated at the center of the trajectory with the electric field rising linearly in a symmetric fashion for displacements from the center. The labels reflect the maximum electric field the ions experience at the outer points of the trajectory. As can be seen in the figure, the ions respond nearly adiabatically to the local micromotion compensation at a given point along the trajectory, making the energy resolution of the ions during MOT interaction less sensitive to the energy resolution of the ions at other points in the trajectory. Approximate regions of MOT interaction are shaded in red (color).}
\label{FigAppendix}
\end{figure}

An uncompensated offset electric field will displace the ion from the rf trap null and lead to excess micromotion energy. Currently, in our system, excess micromotion energy is compensated by iteratively changing the Mathieu-$q$ and Mathieu-$a$ parameter of our trap and adjusting compensation voltages on the LQT electrodes until the ion position changes minimally with trap parameter modulation, as verified through camera imaging. Ultimately our technique is limited by our imaging resolution of $\approx 5$ $\mu$m, leading to excess micromotion uncertainties of $\approx30$ mK. This value could certainly be improved with more sophisticated micromotion-compensation techniques, such as photon correlation spectroscopy or parametric excitation, that have been shown to limit displacements from the null to $\approx 1$ $\mu$m~\cite{Berkeland1998, Keller2015}, which would potentially result in roughly a factor of $25$ improvement in the minimum excess micromotion energy obtainable in our system.

To illustrate how excess micromotion would affect the shuttling trajectory, we performed additional simulations with a single ion being shuttled at low energy, where micromotion more significantly affects energy resolution. In the simulations, stray voltages were imposed on trap electrodes to cause a constant $\approx5$ $\mu$m shift of the ion from the trap null during shuttling (approximately equal to our current micromotion compensation limits), and the results were compared to a perfectly compensated case. The trap and shuttling parameters utilized during these simulations are as follows: $V_{DC}$ $=5$ V, $V_{amp}$ $=1.2$ V, $\gamma$ $=0.03$, and $\omega\sim$ $2 \pi \cdot (25-45)$ Hz.

As can be seen in Fig.~\ref{FigAppendix}(a) and (b), the additional micromotion energy leads both to an upward shift in average kinetic energy as compared to the compensated case and a broader energy distribution. Under perfect compensation, kinetic energy spreads of $\lesssim10$ mK are obtainable, whereas this number is increased to $\lesssim100$ mK in the simulations with excess micromotion. 

Due to stray fields and patch potentials on the electrodes, the offset electric field experienced by the ion may not always be constant and may change during the shuttling trajectory. However, the closest sources of charge that could produce such fields reside on the trap rods, and thus these fields are expected to scale as $\sim1/(r_0-r_e)^2$, where $r_e$ is the electrode radius of our trap rods. While the charges could be arranged in a cluster of any size, point sources may be particularly problematic since without neighboring charges to broaden their potential, the electric field they produce may change significantly along the shuttling trajectory. 

Given the large dimensions of the trap relative to the shuttle trajectory, we can approximate the potential of these fields as 
\begin{equation}
\Phi(r,z) = \frac{Q}{4 \pi \epsilon_0} \frac{1}{(r^2+z^2)^{1/2}}\approx\frac{Q}{4 \pi \epsilon_0}\frac{1}{r}\left(1-\frac{1}{2}\left(\frac{z}{r}\right)^2\right)
\end{equation}

which leads to an radial electric field gradient with respect to the trap axial dimension along the trap null of

\begin{equation}
\begin{aligned}
\frac{\partial}{\partial z}E_r(r,z)|_{r\rightarrow \left(r_0-r_e\right)}&=\frac{\partial}{\partial z}\frac{\partial}{\partial r} \Phi(r,z)|_{r\rightarrow \left(r_0-r_e\right)} \\
&\approx  \frac{Q}{4 \pi \epsilon_0} \frac{z}{\left(r_0-r_e\right)^4}
\end{aligned}
\end{equation} 

where $\Phi(r,z)$ is the electrostatic potential due to a point charge on a rod located a trap electrode, $Q$ is the charge of the point charge of interest, $\epsilon_0$ is the vacuum permittivity of free space, $r$ is the radial dimension of the trap, $E_r(r,z)$ is the component of the electric field in the radial direction, and $(r_0-r_e)$ is the distance between the surface of the electrode and the center of the trap ($r_0-r_e=4.1$ mm in our system). 

While the magnitude of the electric field at any given trajectory point can certainly be significant enough to pull the ion off the null, the fields vary approximately linearly as a function of axial distance. Therefore, we investigated the influence of a linearly varying electric field gradient on the shuttling procedure.

Using the same simulation parameters as in the above-mentioned excess micromotion simulations, we initialized an ion with idealized micromotion compensation at the center of the shuttling trajectory. We then also included an offset electric field that varied linearly with axial trap distance, serving to push the ions off the null as they progressed further from the trajectory center point during the shuttle.

As the total charge producing such patch potentials is difficult to estimate, the results for a variety of reasonable electric field strengths are presented in Fig.~\ref{FigAppendix}(c). Even though the ions experience significant excess micromotion at the trajectory endpoints, since the changes in radial displacement from the null occur gradually, the ions can respond nearly adiabatically to the local excess micromotion amplitude at each trajectory position. Therefore when the ions reach the MOT region, where the micromotion has been compensated adequately, they experience very little excess micromotion. For all field gradients explored in Fig.~\ref{FigAppendix}(c), the kinetic energy widths are predicted to be within $10$\% percent of the perfectly compensated value of $14$ mK, assuming $\omega_A=200$ $\mu$m. 

Of course, multiple charge patches could result in a stronger gradient along the axial trap dimension. If the gradient is strong enough, the ions could respond non-adiabatically to sudden changes in radial positions along the shuttling trajectory, inducing radial secular oscillations that could significantly compromise the energy resolution. However, when imaging the ions while stationary at various points along the trap axis, no such electric field profile is observed, and the ions are found to be within $\approx10$ $\mu$m of the trap null at all points along the trajectory for given set of micromotion compensation shimming voltages (while operating the trap at radial secular frequency of $\approx2\pi \cdot 45$ kHz). Additionally, all fields explored in Fig.~\ref{FigAppendix}(c) above $4.2$ V/m produces ion displacements from the null greater than that observed experimentally, further reducing the likelihood that strong electric fields that could compromise ion energy resolution exist in our system. These results indicate that if compensating micromotion throughout the trajectory is technically infeasible for a give experimental setup, the less challenging task of compensating in the narrow region of MOT interaction may provide similar results.

\end{document}